\documentclass[journal,twoside]{IEEEtran}

\usepackage{cite}
\usepackage{amsmath,amssymb,amsfonts}
\usepackage{algorithmic}
\usepackage{graphicx}
\usepackage{textcomp}

\usepackage{hyperref}  
\usepackage{color}
\usepackage{multirow}

\def\ourmdoel{cofe-Net}
\def\LQA{LQA}
\def\RSA{RSA}
\def\etal{\textit{et~al.}}

\def\BibTeX{{\rm B\kern-.05em{\sc i\kern-.025em b}\kern-.08em
		T\kern-.1667em\lower.7ex\hbox{E}\kern-.125emX}}

\begin{document}
	\title{Modeling and Enhancing Low-Quality Retinal Fundus Images}
	
	\author{Ziyi~Shen, Huazhu~Fu, Jianbing~Shen, and Ling~Shao
		\thanks{Z.~Shen, H.~Fu, J.~Shen, and L.~Shao are with Inception Institute of Artificial Intelligence, Abu Dhabi, UAE. (email: \{ziyi.shen, huazhu.fu, jianbing.shen, ling.shao\}@inceptioniai.org)}
		\thanks{Corresponding authors: \textit{Huazhu~Fu}  and  \textit{Jianbing~Shen}. }}

	\maketitle
	
	\begin{abstract}
		Retinal fundus images are widely used for the clinical screening and diagnosis of eye diseases. However, fundus images captured by operators with various levels of experience have a large variation in quality. Low-quality fundus images increase uncertainty in clinical observation and lead to the risk of misdiagnosis. However, due to the special optical beam of fundus imaging and structure of the retina, natural image enhancement methods cannot be utilized directly to address this. In this paper, we first analyze the ophthalmoscope imaging system and simulate a reliable degradation of major inferior-quality factors, including uneven illumination, image blurring, and artifacts. Then, based on the degradation model, a clinically oriented fundus enhancement network (cofe-Net) is proposed to suppress global degradation factors, while simultaneously preserving anatomical retinal structures and pathological characteristics for clinical observation and analysis. Experiments on both synthetic and real images demonstrate that our algorithm effectively corrects low-quality fundus images without losing retinal details. Moreover, we also show that the fundus correction method can benefit medical image analysis applications, e.g., retinal vessel segmentation and optic disc/cup detection.
	\end{abstract}
	

	\section{Introduction}
\begin{figure}[!t]
	\centering
		\includegraphics[width=1\linewidth]{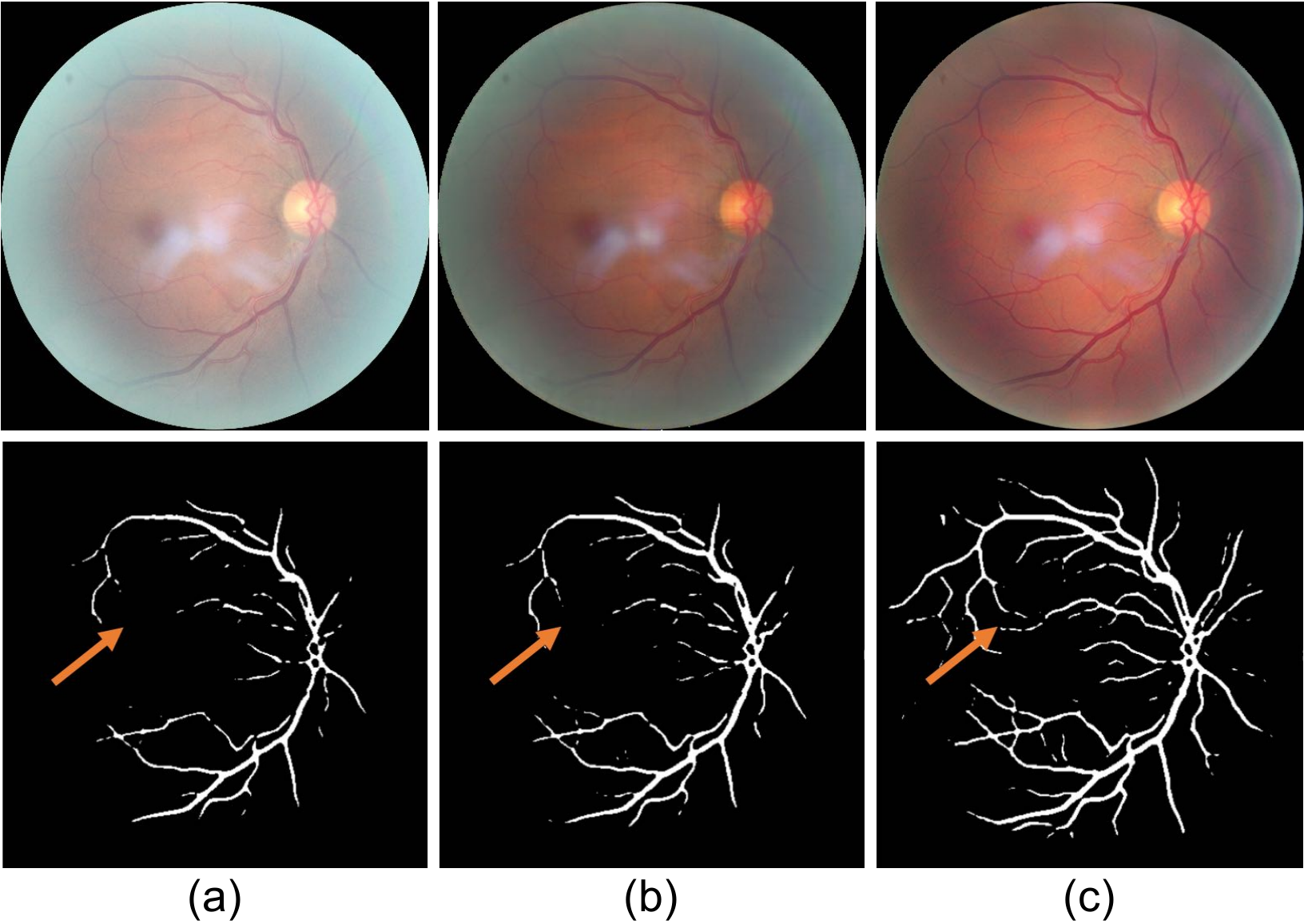} 
		\caption{\textbf{Illustration of fundus image correction.} (a) The low-quality fundus image. (b) Result enhanced by~\cite{EilertsenKDMU17}. (c) Result enhanced by our method. The first row shows the fundus images, and bottom row is the corresponding vessel segmentation results using an automated method~\cite{vessel_seg}. Our method corrects the low-quality fundus image while enhancing the clinical structures (e.g., vessel, optic disc and cup regions).}
		\label{fig:begin}
	\end{figure}

	Due to their safety and cost-effectiveness in acquiring, retinal fundus images are widely used by both ophthalmologists and computer-aided diagnosis systems for the clinical screening and diagnosis of ocular diseases~\cite{Abramoff2010,Schmidt-Erfurth2018}. However, fundus images tend to experience large variations in quality. A screening study of 5,575 patients found that about 12\% of fundus images are not of adequate quality to be readable by ophthalmologists~\cite{Philip891}. In some cases, when the degradation is caused by the images being obtained through  \textit{internal} cataractous turbid media,   enhancement methods, such as~\cite{peli1989}, can be used to restore `high quality'. 
	Then, the corrected images can be used to support the observation of other diseases (e.g., age-related maculopathy, diabetic retinopathy, and glaucoma).
	However, in addition to this pathogenic degradation, in real applications,  \textit{external} interference factors caused by handcrafted imaging equipment and poor environmental conditions are also common.
	For instance, images are often taken under different lighting environments, using various cameras, and by distinct operators with varying levels of experience. 
    Common examples of low-quality factors in retinal fundus images thus include uneven illumination, image blurring, and artifacts, which not only prevent reliable diagnosis by ophthalmologists, but also affect the performance of automated image analyzing systems~\cite{EyeQdata2019,Cheng2020}. 
    An example is shown in Fig.~\ref{fig:begin}~(a), where uneven illumination and artifacts prevent the vessel and disc region from being fully/clearly observed, and affect the performance of the automated vessel segmentation method (i.e.,~\cite{vessel_seg}).
	 
	Recently, general image enhancement methods have achieved state-of-the-art performances, especially with the development of deep learning techniques~\cite{YangXS0WL18,lowlight1}. However, different from general images, retinal fundus  images are acquired through a special ophthalmoscope imaging process to capture  anatomical retinal structure for clinical diagnosis, which introduces various additional challenges. 
	First, the retina cannot be illuminated internally; both the incident and reflected imaging beams have to traverse the pupil.  
	Moreover, the spherical geometry of the eye creates significant inter-reflection, resulting in shading artifacts~\cite{FORACCHIA2005179}.  
	Second, anatomical retinal structures (e.g., vessel, optic disc and cup) in fundus images are limited but highly important for clinical diagnosis, and should thus be enhanced in the correcting process.
	Third, some pathological characteristics (e.g., hemorrhages, microaneurysms, and drusen) are usually only a few pixels wide and appear as circular shapes, causing them to be easily confused with artifacts and noise.
	These issues mean that fundus image correction methods need to be able to both suppress the undesired low-quality factors and preserve the pathological characteristics simultaneously, which general enhancement techniques cannot do to a satisfactory level. For example, Fig.~\ref{fig:begin}~(b) shows the enhanced result of Fig.~\ref{fig:begin}~(a) when using the general image enhancement method~\cite{EilertsenKDMU17}, where the disc region still suffers from artifacts, and vessels are miss-segmented by the automated system.

	To address these issues, in this paper, we design a degradation model that simulates major factors of low-quality fundus images, including light transmission disturbance, image blurring, and retinal artifacts. Then,  a clinically oriented fundus enhancement method~(\ourmdoel)  is proposed to suppress the local outliers and undesired artifacts, while at the same time preserving the anatomical retinal structures, e.g., vessel and optic disc/cup regions. To this end, two new modules, the retinal structure activation (\RSA) and clinical low-quality activation (\LQA),  are introduced. The \RSA~module is used to preserve the retinal structure, while the~\LQA~is employed to remove the low-quality factors. 
	Based on the human perception mechanism, the proposed network with additional error metrics for perceiving artifacts, is able to correct fundus images with more accurate structures and suppress local defects. An example of a fundus image corrected by our method is shown in Fig.~\ref{fig:begin}~(c).
	The main contributions of this paper are summarized as follows:
	\begin{enumerate}
		\item A fundus degradation model based on the retinal ophthalmoscope imaging system is designed to simulate low-quality fundus images. It can be widely utilized to support the typical propagation scheme in fundus image generation models. To the best of our knowledge, this is the first work to model the optical ophthalmoscope. All degradation models are designed based on several imaging stages.
		\item A novel clinically oriented fundus enhancement network~(\ourmdoel) is developed to correct low-quality fundus images for clinical observation and analysis. Our~\ourmdoel~preserves the anatomical retinal structures of fundus image using the \RSA~module and suppresses  undesired artifacts with the \LQA~module. 
		\item We show that fundus correction can boost the performances of clinical analysis tasks, \textit{e.g.},  vessel segmentation and disc/cup detection, on poor-quality images. Experimental results on both synthetic and real fundus images demonstrate that our algorithm performs favorably against state-of-the-art approaches\footnote{The source code of our method is available at \textcolor{red}{\url{https://github.com/HzFu/EyeQ_enhancement}}}.
	\end{enumerate}

	\section{Related Work}
	In this section, we summarize current image correction techniques and discuss the algorithms that have been specifically applied to fundus images. 
	Several methods~\cite{ForacchiaGR05,hwang2012context} exploit image contrast normalization and contrast limited adaptive histogram equalization (CLAHE) techniques to restore an image.
	For example, Setiawan~\etal~\cite{retinaclahe} applied CLAHE to fundus image enhancement specifically.
	Instead of simply considering the color and texture information, some algorithms~\cite{ng2011total,wang2014nonlocal,fu2016weighted} decompose the reflection and illumination, achieving image enhancement and correction by estimating the solution in an alternate minimization scheme. 
	Guo~\etal~\cite{lime} further proposed to refine the illumination map and enhance low-light images. 
	These models have been extended to an integrated scheme incorporating gamma correction~\cite{huang2012} and CLAHE for fundus image luminance and contrast adaption~\cite{zhou2017color,naturalness2}. 
	While these algorithms based on a bottom-up framework are effective, their optimal solution relies heavily on global image statistics and mapping functions, ignoring discriminative features, which may introduce undesired artifacts and distortion.
	
	Along another line, learning based methods have also been developed for the image correction task, utilizing the various features extracted from the images to learn directional filters. 
	For instance, latent image priors have been adopted for correction and restoration in sparsity-based models~\cite{osher2005,LuYY13}, distribution fitting algorithms~\cite{Daniel2011,LevinWDF09,convex2}, variational frameworks~\cite{convex3}, and latent structure-driven methods~\cite{RenCPGZY16,He0T11,naturalness1}, all of which have also been specifically applied to fundus images~\cite{vessek_enh1,vessek_enh2,BaghaieDY15,CHENG}. 
	These algorithms typically constrain the optimal solution using a regularization scheme to solve the non-convex problem.        
	This may incur a heavy computational cost, limiting their applicability in clinical settings.  
	
	Recently, due to the powerful image representation ability, deep learning techniques have been widely used in computer vision and medical image analysis.
	This has enabled the rapid advancement of reconstruction techniques, making them much better equipped to address various challenging tasks, such as low-light image enhancement~\cite{lowlight1,lowlight2,wei2019single}, dehazing and deraining~\cite{haze,rain,Li2020_TMM}, and deblurring~\cite{deblur1,deblur2}.
	For image correction, convolutional neural network~(CNN) based approaches attempt to learn a mapping operator between the ground truth and low-quality images~\cite{EilertsenKDMU17,RenLZPC016,LvLWL18,Li2020_ACMMM}. 
	Eilertsen~\etal~\cite{EilertsenKDMU17} proposed to solve the high dynamic range task in an end-to-end fashion, under the constraint of a pixel-wise loss. This method produces accurate global tone-mapping but results in an over-smooth solution. 
	Ren~\etal~\cite{RenLZPC016} and Lv~\etal~\cite{LvLWL18} aimed to enhance images in a coarse-to-fine manner, using a multi-scale framework and a feature fusion mechanism, respectively. 
	In addition, Talebi and Milanfar~\cite{TalebiM18} introduced a deep neural image assessment model and applied it to restore more content under extreme conditions (e.g., dark and bright areas).   
	These end-to-end methods aim to learn an optimal solution by simply minimizing the content loss. 
	Focusing on images stocked with explicit prior information, Liu~\etal~\cite{LiuMWZ19} proposed a deep prior ensemble and integrated knowledge-driven cues for natural image enhancement.
	Different from the non-convex optimization framework, deep learning methods follow a heuristic pattern that relies heavily on a great deal of training data. 
	Due to the particular pathological characteristics and disease markers, general CNN-based models often do not perform well on medical samples, especially fundus images.  
	To deblur and enhance clinical images, Zhao~\etal~\cite{ZhaoYCL19} and Liu~\etal~\cite{UpadhyayA19} applied the adversarial loss. 
	However, while computationally efficient, these methods only focus on generating photo-realistic images, ignoring the lesion areas significant to clinical applications.
	Therefore, designing an effective deep learning model for fundus image correction is the focus of this work. 
	
	\begin{figure}[t]
		\centering
		\includegraphics[width=0.9\linewidth]{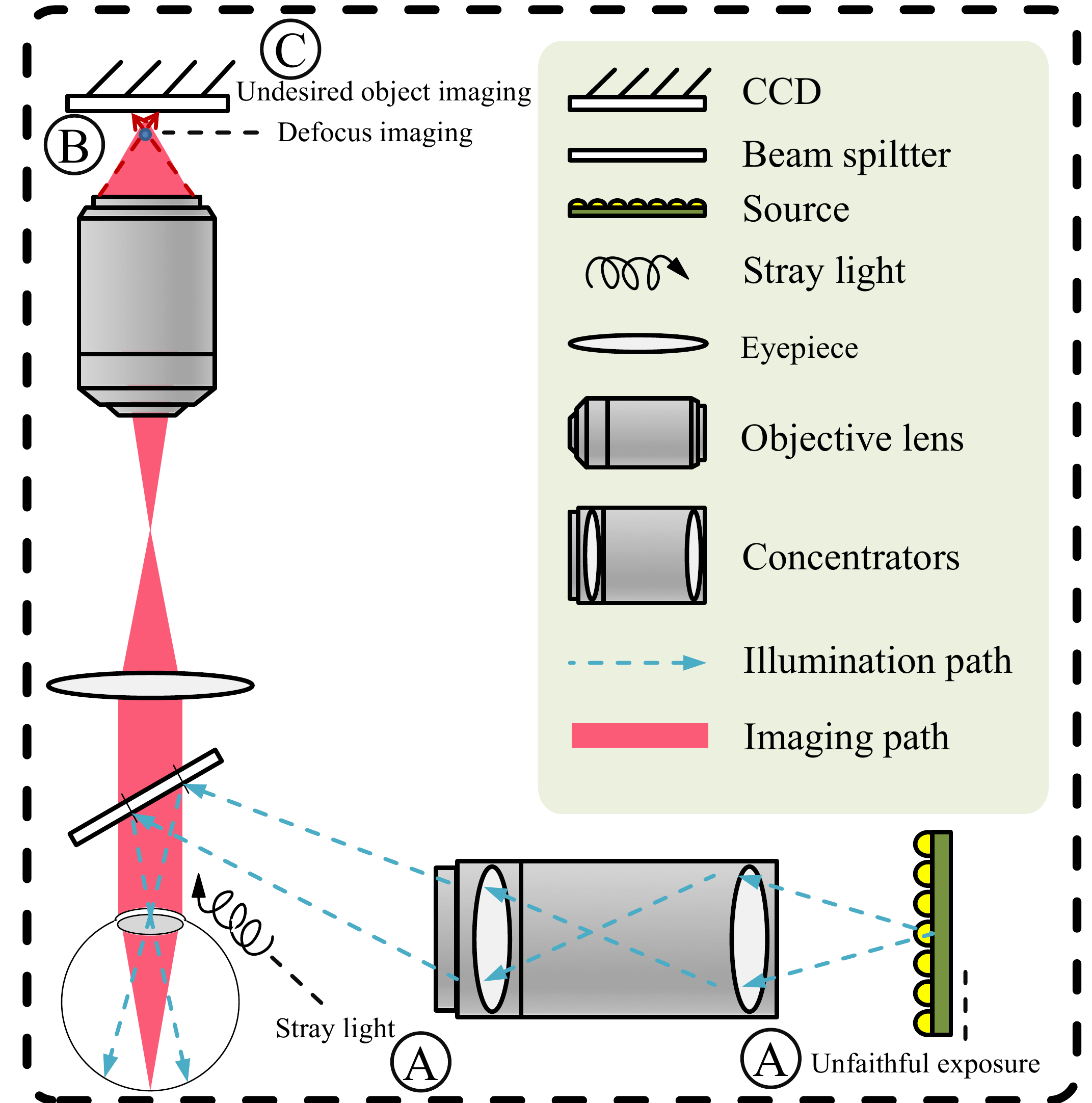}
		\caption{\textbf{Schematic diagram of ophthalmoscope imaging system.} It contains an illumination system and an imaging system. The clinical imaging process in practice undergoes several types of interference, such as exposure and imaging outliers, which introduce unexpected drawbacks to the retinal fundus images and result in image degradation. Besides, the capturing of undesired objects is also a crucial factor that interferes with the image quality and impacts the diagnostic process.}
		\label{fig:system}
	\end{figure}

	\section{Fundus Image Degradation Model} 
	
	Clinical image collection in a complex environment using an ophthalmoscope often encounters several types of interference, introduced in the optical feed-forward system. 
	For instance, as shown in Fig.~\ref{fig:system}, light transmission disturbance is often caused by exposure issues. 
	Due to the interspace between the eye and camera, stray light may enter into the ophthalmoscope, mix with the lighting source and result in uneven exposure. 
	This also affects the tuning setting of the programmed exposure,  causing global over-/under-exposure. 
	In addition, image blurring caused by human factors (such as eyeball movement, fluttering, and defocus) results in low-quality images. 
	Besides, the capturing of undesired objects (e.g., dust) during imaging is also a crucial factor that reduces image quality and impedes subsequent diagnosis.   
	In this section, we propose a reformulated representation of the interference that occurs during the collection of fundus images.  Our degradation model could be used to not only support current fundus propagation models, but also synthesize a high-quality pairwise fundus dataset for subsequent research.  
	We summarize the interference in terms of three factors, including \emph{light transmission disturbance},  \emph{image blurring}, and  \emph{retinal artifacts}. 
	 
	\begin{figure}[!t]
		\centering
		\includegraphics[width=1\linewidth]{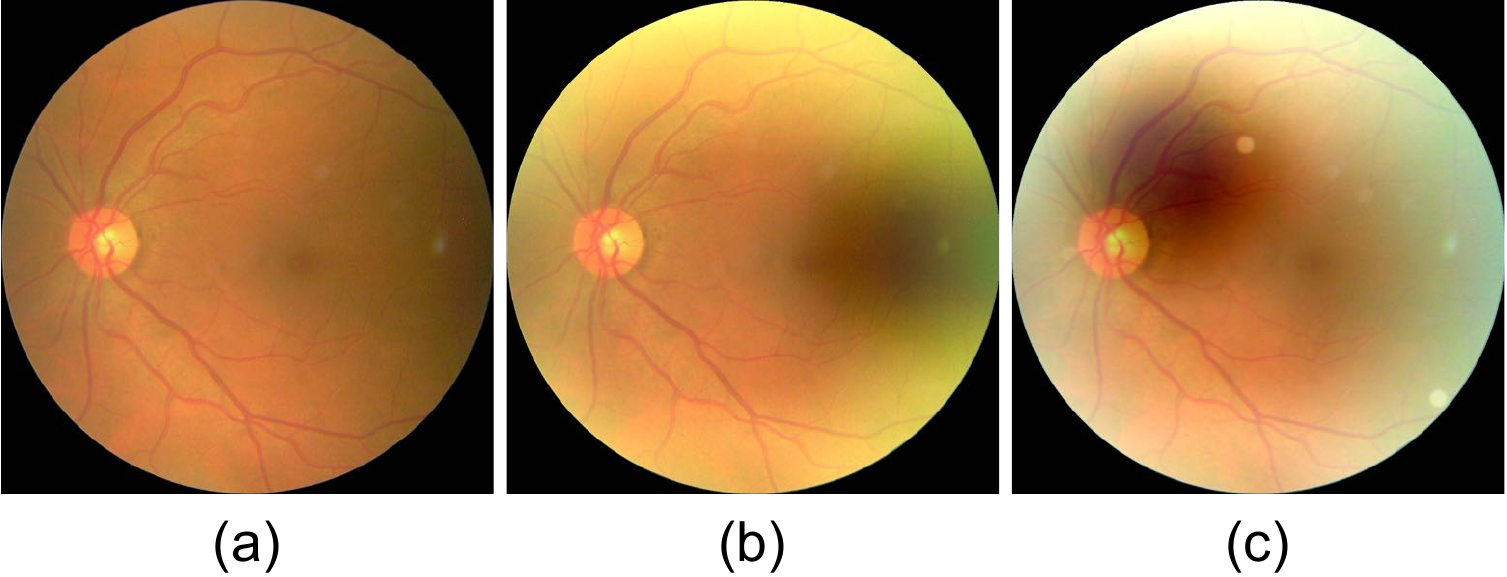}
		\caption{\textbf{Synthetic degraded fundus images.} (a) Clear image. (b) Simulated sample undergoing light transmission disturbance. (c) Simulated sample undergoing light transmission disturbance, image blurring, and retinal artifacts. }\label{fig:sample}
	\end{figure}
	
	\noindent{\textbf{Light Transmission Disturbance:}}
	We first introduce the interference caused by light, which can be categorized into two types, global and local factors. 
	For global factor, since existing fundus cameras are programmed with auto-exposure, ambient light will impact the illumination. 
	Unstable stray light, which affects the camera configuration (light source), may result in under-/over-exposure during image collection.   
	Furthermore, the subjective situation and manual mydriasis also cause disturbance during the imaging procedure.  
	For local factor, the sensitivities of specific regions on the image plane may differ, causing uneven illumination in an image.
	This can be due to the interspace between the fundus and ophthalmoscope. An initiative light leak phenomenon caused by the patient or an inappropriate exposure imported to the imaging plane with a variant distance to the optical axis will result in uneven illumination. 
	In addition, depending on the equipment design of the ophthalmoscope system, the diverse lens apertures and embedded optical compensation mechanism can also limit the amount of light, as well as affect the dynamic range of the fundus images.
	Here, we simulate the degradation using an aggregation model.
	The global factors are modeled by contrast, brightness and saturation interference, while the local factors are defined as additional non-uniform illumination on the fundus image.
	To formulate the light degradation model, given a ground truth (clean) image $\mathbf{x}$,  its degraded image $\mathbf{x}'$ with light transmission disturbance is defined as: 
	\begin{equation}
	\mathbf{x}' = \text{clip}\left(\alpha( \mathbf{J} \cdot G_L\left( r_L,\sigma_L  \right)  + \mathbf{x})+\beta; s\right),
	\label{eq_d_light}
	\end{equation}
	where $\alpha$, $\beta$ and $s$ refer to the contrast factor, brightness and saturation, respectively. 
	We use a clipping function $\text{clip}\left(\cdot;s\right)$ to model the global degradation as a saturation process. 
	For local light leak/lack, $\mathbf{J}$ is defined as an illumination bias to be over-/under-illuminated at a panel centered at $(a,b)$ with a radius of $r_L$.  Therefore, each of its entries is defined as: 
	\begin{equation}
	    \mathbf{J}_{ij}=n_l|_{(i-a)^2+(j-b)^2<r_L^2},
	\end{equation}
	where $n_l$ is the over-/under-intensity weight.
	A Gaussian kernel $G_L$ is then applied to ensure luminance smoothness.
	Fig.~\ref{fig:sample}~(b) gives an example with over-exposure and uneven illumination synthesized by our degradation model.
	%
	
	\begin{figure*}[!t]
		\centering
		\includegraphics[width=1\textwidth]{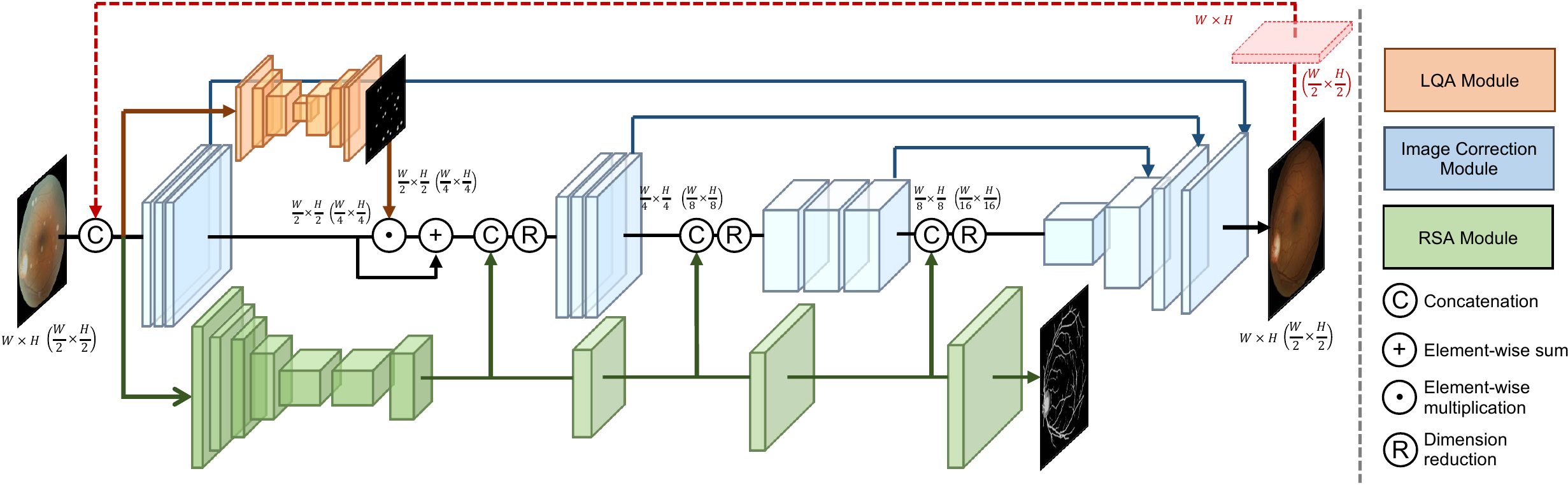} 
		\caption{\textbf{Architecture of the proposed network.}  A low-quality image is simultaneously rendered to three sub-modules, i.e., the low quality activation (\LQA) module, the retinal structure activation (\RSA) module and the image correction module. The \LQA~module activates and embeds the representation of the low-quality region into the correction module at an early stage, while the \RSA~module continuously injects retinal structural information to the correction network by feature map concatenation at each scale. Leveraging this additional information, the image correction module enhances high-quality images based on an encoder-decoder network. Note that we cascade the network for two scales ($0.5 \times$ and $1 \times$) using shared weights. The red dotted line denotes the upsampled result of $0.5 \times$ scale, which is only used in the first $0.5 \times$ loop.
		} \label{fig:Framework}
	\end{figure*}
	
	\noindent{\textbf{Image Blurring:}}
	During the fundus imaging procedure, in addition to issues caused by the program settings, human operator error can also be introduced. 
	Since the ophthalmoscope is applied externally on patients, the distance between the fundus and object plane is a random variable;
	a wrong setting for the focal length of the optical system during the funduscopy, or a dilated pupil caused by light simulation, can cause an undesired object distance between the image plane and lens, and result in image blurring. 
	To simulate this, we define the defocus blurring as: 
	\begin{equation}
	\mathbf{x}'=\mathbf{x} \cdot G_B\left ( r_B,\sigma_B  \right )+n,
	\label{eq_d_blur}
	\end{equation} 
	where $G_B$ is a Gaussian filter with a radius $r_B$ and spatial constant $\sigma_B$, and $n$ denotes the additive random  Gaussian noise. The degraded fundus image $\mathbf{x}'$ is generated via a convolution operation between the ground truth image $\mathbf{x}$ and a Gaussian filter. Specifically, the blur radius $r_B$ is defined as: 
	\begin{equation}
	r_B =\frac{F \cdot \nu_{0}-D\left(F-\nu_{0}\right)}{D \cdot f},
	\label{eq:image} 
	\end{equation} 
	where $F$ is the focal length of the optical lens system, and $f$ is its f-number. $D$ and $\nu_{0}$ denote the object distance and image distance, respectively. 
	
	\noindent{\textbf{Retinal Artifact:}}
	The above formulated models simulate internal factors, which are caused by the imaging system itself. 
	However, additional degradation may be introduced if the imaging is conducted under a poor condition. 
	Although not additive or multiplicative noise, dust and grains attaching on the lens of the imaging plane can also yield blurred images that affect the fundus image quality and following diagnosis.
	We define this type of interference as retinal artifact, which is modeled by a multi-step imaging procedure:
	\begin{equation}
	\mathbf{x}^{\prime}=\mathbf{x}+\sum_{k}^K G_R\left(r_{k} / 4, \sigma_{k}\right) \cdot \mathbf{o}_{k}.
	\label{eq_d_art}
	\end{equation}
	In the context of random light fields, $r_k$ and $\sigma_{k}$ are defined as the radius and variance of an undesired object $k$ imaged on the plane, where $r_k$ can be calculated as the same way in Eq.~\ref{eq:image}. $K$ is the undesired object number. Parameter $\mathbf{o}_k$ is the luminance bias.
	We again define a Gaussian filter $G_R$ to simulate the defocused imaging of an undesired object.
	
	\noindent{\textbf{Degradation:}} Totally, we propose three different degradation models in Eqs.~(\ref{eq_d_light}, \ref{eq_d_blur}, and \ref{eq_d_art}), which simulate  light transmission disturbance, image blurring, and retinal artifacts, respectively. These models can be utilized independently to synthesize individual low-quality factor, or randomly combined for the more complex degradation. An example containing various types of degradation is shown in  Fig.~\ref{fig:sample}~(c), where our proposed degradation model is particularly well-designed to simulate the real funduscopic examination procedure.  
	
	\section{Clinically Oriented Fundus Enhancement Network} 
	
	For medical image analysis, there are certain crucial cues that require special attention. 
	Thus, not only is it necessary to preserve the fine structures in a clinical image, but also the pathological characteristics. 
	In other words, it is essential to correct the images while simultaneously guaranteeing they remain clinically significant for disease diagnosis and analysis. 
	Therefore, we propose a clinically oriented fundus enhancement network~(\ourmdoel) to correct the low-quality fundus image, as shown in Fig.~\ref{fig:Framework}. Given a low-quality fundus image, a low-quality activation (\LQA) module is employed to identify the retinal artifacts, while a retinal structure activation (\RSA) module is used to perceive the latent retinal structure. Then, an image correction module is introduced to fuse these prior information with the latent features of the image to produce the final corrected image.

	\subsection{Low-Quality Activation Module} 
	
	To enhance fundus images and encourage valid and accurate pathological analysis of ocular diseases, it is necessary to exploit a learning-based solution that extracts latent features from the low-quality fundus image and learns a good representation to  reconstruct it. 
	Thus, we first define a generative network $\mathcal{G}$. 
	Let $\mathbf{I}$ and $\hat{\mathbf{I}}$ denote the low-quality input image and the corrected image, respectively, with size $w \times h \times c$, where $w$ and $h$ denote the width and height of the image, and $c$ is the channel number (e.g., $c=3$ for an RGB image). 
	The correction procedure can be formulated as: 
	\begin{equation}
	\hat{\mathbf{I}}=\mathcal{G}\left(\mathbf{I} ; \mathbf{W}_{\mathcal{G}}\right), 
	\end{equation} 
	where $\mathbf{W}_{\mathcal{G}}$ denotes the learnable parameters of network $\mathcal{G}$.
	To progressively generate the corrected images, a robust $\mathcal{L}_2$ loss is used as the object function:
	\begin{equation}
	\mathcal{L}_{c}=\| \mathbf{I}_G -\mathcal{G}\left(\mathbf{I}, \mathbf{W}_{\mathcal{G}}\right)\|_{2}^{2},
	\end{equation}
	where $\mathbf{I}_G$ is the ground truth (clean) image.
	However, only using the $\mathcal{L}_2$ loss to guide the convergence of the neural network often results in an over-smooth solution due to the similarity between the input image and the restored image in terms of their pixel-wise appearances.
	Therefore, the vanilla model simply pays equal attention to all areas of the global image, without taking clinical regions (e.g., vessels, disc/cup regions) into special consideration. The model may thus fail to remove local artifacts, or preserve real structures.
	
	To address this issue, we introduce a supervised activation mechanism, the \LQA~module, to perceive the undesired low-quality factors in clinical images.
	Our \LQA~module $\mathcal{M}$ is built upon a convolutional encoder-decoder   architecture, as shown in red part of Fig.~\ref{fig:Framework}, and produces artifact attention map $\mathbf{m}$ corresponding the low-quality region. This differentiable neural activation operator can be defined as:
	\begin{equation}
	\mathbf{m}=\mathcal{\sigma}(\mathcal{M}(\mathbf{I},\mathbf{W}_{\mathcal{M}})),
	\end{equation}       
	where $\mathcal{\sigma}$ denotes the non-linear activation function that normalizes the feature. 
	The proposed correction framework is designed to perceive local artifacts via  \LQA~branch with parameters $\mathbf{W}_{\mathcal{M}}$, and embed the corresponding map to guide the correction. 
	Let $\mathbf{F}_\mathcal{L}$ denote the low-level features extracted by the image correction network. 
	We combine the artifact attention map $\mathbf{m}$ of \LQA~and $\mathbf{F}_\mathcal{L}$ as: $\mathbf{F}_\mathcal{L} \odot \mathbf{m} + \mathbf{F}_\mathcal{L}$, where $\odot$ is an element-wise multiplication.
	The artifact aware features extracted from the \LQA~are integrated with the low-level features into the correction model.         
	Therefore, the embedded framework encourages a more efficient representation and achieves a higher ability for distinguishing the low-quality region.
	Here, we utilize a pixel-wise $\mathcal{L}_2$ loss to constrain the \LQA~module training:
	\begin{equation}
	\mathcal{L}_{m}=\|\mathbf{m_G}-\mathbf{m} \|_{2}^{2},
	\end{equation}   
	where $\mathbf{m_G}$ denotes the ground truth artifact mask.
	The supervision enables the \LQA~branch to efficiently learn an activation representation.
	Beside the $\mathcal{L}_2$ loss function, we also apply another relevant loss function to identify the undesired artifacts and train the correction network by minimizing the low-quality perceiving loss: 
	\begin{equation}
	\mathcal{L}_{p}=\| \mathbf{m} \odot
	 (\mathbf{I}_G -\mathcal{G}\left(\mathbf{I} ; \mathbf{W}_{\mathcal{G}}\right))\|_{2}^{2},
	\end{equation}    
	where $\mathbf{I}_G$ is the  the ground truth (clean) image. This low-quality perceiving loss encourages to highlight the correction of artifact regions.
	The \LQA~module can assist the correction by focusing on the undesired artifacts.
	
	In summary, the proposed \LQA~module aims to highlight the representation of retinal artifact in the fundus image. 
	Since undesired objects (e.g., dust and grains) on the object lens often have similar appearances, we tend to merely perceive the shape and position of this kind of artifact. 
	Therefore, we apply a small-scale network for the \LQA~module and set the input to be a $0.5 \times$ image to reduce the computational load. 
	As shown in Fig.~\ref{fig:Framework}, the \LQA~module contains three encoder for extracting the features, and the perceived information is integrated and represented by the three corresponding encoder blocks. 
	Each encoder block contains three convolutional layers with $3\times 3$ kernels followed by a $2 \times$ max-pooling layer for downsampling, while each decoder block includes one $3\times 3$ convolutional layer and one $4\times 4$ transposed convolutional layer for upsampling the feature maps. Each convolutional layer  is followed by a ReLU function as the non-linear activation. 
	Finally, a $1 \times 1$ convolutional layer is utilized to predict the artifact attention map.
	
	\subsection{Retinal Structure Activation Module} 
	
	Natural image enhancement method tends to reconstruct images with more diverse context in order to please human perception, often containing exaggerated details. 
	However, in some cases, this may not have positive impact. This is especially true for medical images, such as fundus images, which are used as the foundations for clinical diagnosis. The incorrectly synthetic content may heavily skew the pathological description. 
	
	To address this issue, we introduce a retinal structure activation (\RSA)~module to preserve features of the retinal structure and guide the reconstruction processing. 
	As illustrated in Fig.~\ref{fig:Framework}, the \RSA~module first encodes the input image using a pre-trained ResNet-34. 
	The features are directly extracted by the decoding operator for the following structure extraction. 
	Then, the decoding residual layers are used to scale up the bottleneck features (by a factor of 16) to the original image size. 
	An additive skip connection from the encoder is applied to the corresponding decoding layers.  Our \RSA~module predicts a segmentation mask map for main retinal structures, e.g., vessel, optic disc and cup regions. We utilize a mean squared error (MSE) loss as the segmentation loss to constrain the prediction and simultaneously, the non-linear \RSA~feature maps obtained from different depths are explicitly utilized to assist the main correction model. This is formulated as: 
	\begin{equation}
	\mathbf{S}^{l}= \mathcal{T}_\mathcal{S}^{n}\left(\text{Concat}\left[\mathbf{F}_\mathcal{S}^{n}\left(\mathbf{I}\right), \mathbf{F}_\mathcal{G}^{n}\left(\mathbf{I}\right)\right]\right), 
	\end{equation}
	where $\mathbf{F}_\mathcal{S}^{n}\left(\mathbf{I}\right)$ and $\mathbf{F}_\mathcal{G}^{n}\left(\mathbf{I}\right)$ denote the features extracted from the $n^{th}$-level of the \RSA~module and image correction network, respectively. $\text{Concat}\left[\cdot , \cdot \right]$ is a concatenation operator, and the \RSA~map is fused via a learnable non-linear transformation filter $\mathcal{T}_{S}^{n}\left(. \right)$. 
	The feature maps from different levels are stacked and embedded with directional information to enable an accurate mapping between the features of the \RSA~module and our image correction module. 
	Thus, they are able to provide the network with a stronger representational ability, and preserve more realistic and meaningful  content. 
	In this way, the key components of the fundus image are encoded in the hidden states of the image correction network, considerably reducing the problems of content vanishing and singular positioning.

	\subsection{Image Correction Module}
 
	With the assistance of the \LQA~and \RSA~modules, we design the image correction network $\mathcal{G}$ as a two-scale cascaded framework  to progressively enhance images in a coarse-to-fine manner. Specifically, we propose to train a single encoder-decoder network for two scales ($0.5 \times$ and $1 \times$) by sharing weights to reduce the number of trainable parameters. 
	As shown in Fig.~\ref{fig:Framework}, we first encode the image, inferring features by identifying the critical information. This is done by densely embedding (concatenating) the knowledge learned from the \RSA~module. The features aggregated at each level are integrated and their dimensions are reduced by a convolutional layer. Then, a symmetric decoder is applied to generate the corrected image. We also add skip connections between the encoder and corresponding decoder to avoid gradient explosion or vanishing during training. 
	For the two-scale cascaded framework, we first feed the input image with a downsampled size (i.e., $0.5 \times$ scale) to the network, and then utilize a transposed convolutional layer to produce the corrected result with the original size (i.e., $1 \times$ scale). After that, the input image is concatenated with the upsampled corrected result and fed into the image correction network again to be processed at the $1 \times$ scale. Note that the input of the $1 \times$ scale is comprised of the upsampled result from the $0.5 \times$ scale and the original image, while the input of our image correction network has six channels. We thus duplicate the input image of the first scale (i.e., $0.5 \times$ scale) to provide the same number of channels.
	Finally, the overall loss function for our whole network is defined as: 
	\begin{equation}
	\mathcal{L}=\mathcal{L}_{m} +\sum_{s} \left(\lambda_{p}\mathcal{L}_{p}^{s} + \lambda_{c}\mathcal{L}_{c}^{s}\right) +\lambda_{v}\mathcal{L}_{v}, 
	\end{equation}
	where $\mathcal{L}_{v}$ is the MSE loss of the \RSA~module, and $s$ denotes the scale index (e.g., $s = 0.5, 1$). $\lambda_{p}$, $\lambda_{c}$ and $\lambda_{v}$ are weights that balance the different losses. Note that $\mathcal{L}_{p}$ and $\mathcal{L}_{c}$ are calculated at both scales to control the image reconstruction. $\mathcal{L}_{v}$ and $\mathcal{L}_{m}$ are simultaneously computed only at the original scale (i.e., $1 \times$). Here we set  $\lambda_{p}=10$, $\lambda_{c}=1$,  $\lambda_{v}=0.1$. These penalty terms are employed to stabilize the training process and ensure the corrected image is visually reasonable. 
	
	\section{Experiments}\label{exp}
	
	\subsection{Datasets}
	To train our \ourmdoel, we manually select 13,000 high-quality images, free from interference factors, from the EyeQ dataset~\cite{EyeQdata2019}, which is based on the Kaggle Diabetic Retinopathy Detection dataset~\cite{kaggle}.
	We then randomly choose degradation factors (e.g., light transmission disturbance, image blurring, and retinal artifacts), and process the images using the proposed degradation formulation to generate low-quality images, which make up the training set. 
	Note that we also process the images by combining all factors to simulate a complicated real-world situation.     
	For testing,  we also utilize the proposed degradation model to randomly generate degraded images for quantitative evaluation.
	We synthesize 500 testing images in total using the Kaggle~\cite{kaggle} dataset and DRIVE~\cite{staal2004ridge} dataset.
	Here peak signal to noise ratio~(PSNR) and structural similarity~(SSIM) are utilized to evaluate the image quality.
	In addition, for qualitative analysis, we randomly choose 50 images from the Kaggle dataset~\cite{kaggle}, and build a benchmark to support a user study task.
	
	\begin{figure}[!t]
	\centering
	\includegraphics[width=1\linewidth]{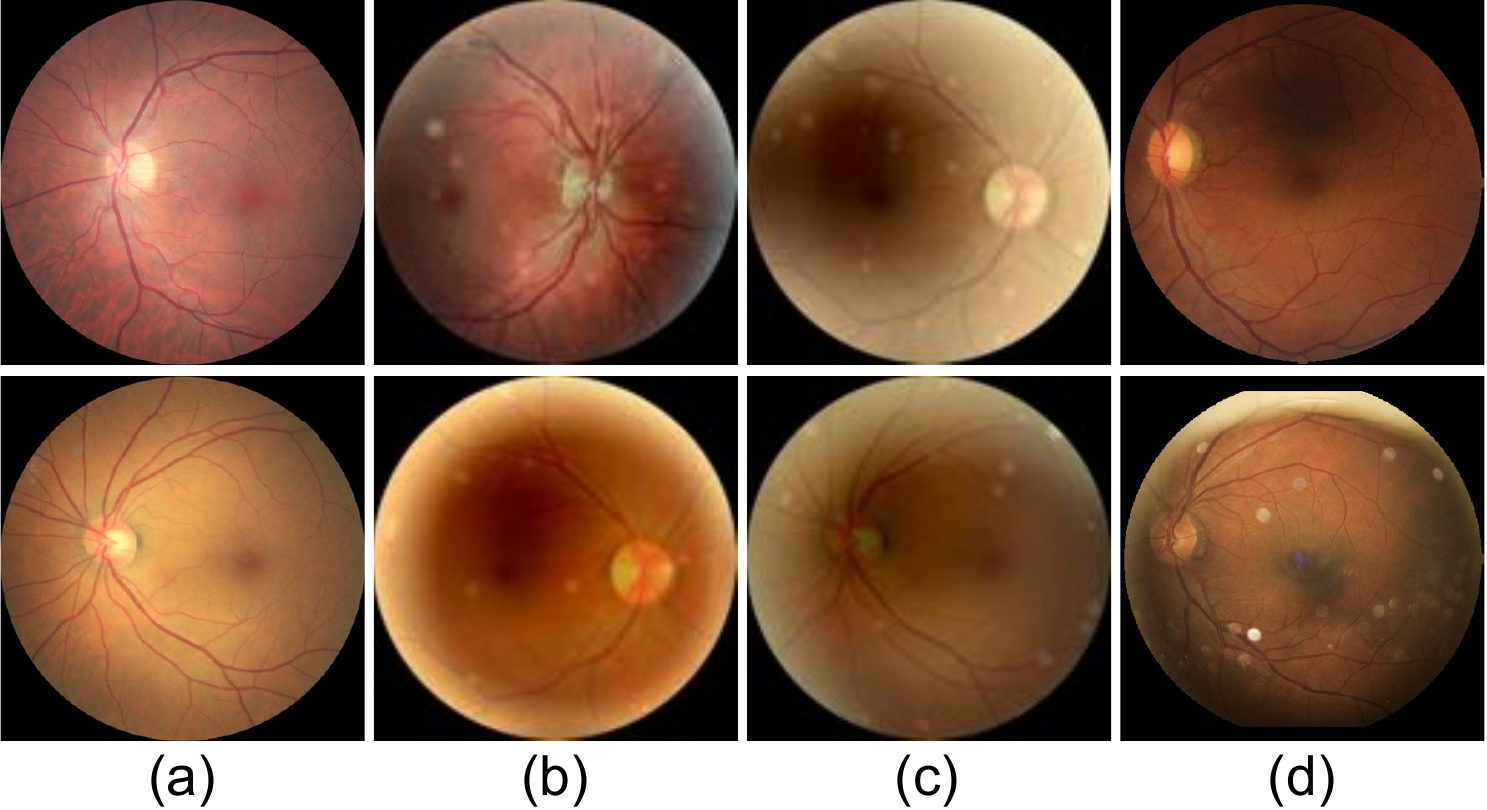}
	\caption{\textbf{Synthetic samples in low-quality fundus image dataset.} 
		(a) Ground truth images. (b-c) Synthesized low-quality images. (d) Real low-quality images.
	}\label{fig:samples}
	\end{figure}

	\subsection{Implementation}
	
	During training, we resize the degraded input images to 512$\times$512 with a batch size of 16 for each iteration. 
	To accelerate the training process, we first train the \RSA~and \LQA~modules, and then apply an end-to-end training strategy to the whole correction framework.  
	The proposed \ourmdoel~is implemented in PyTorch, using stochastic gradient descent~(SGD) for optimization. 
	The learning rate is initialized to $1\times10^{-4}$ for the first 150 epochs, and then gradually decayed to zero over the next 150 epochs. 
	The network is trained on an NVIDIA Tesla V100 GPU with 32GB of memory per card.  
	The number of model parameters is roughly 41M, and the computational cost is 270~GMac.
	 
	\subsection{Degradation Model Settings}
	In the proposed retinal fundus image degradation model, we summarize the interference in terms of three factors, including light transmission disturbance, image blurring and retinal artifacts. 
	To better simulate real clinical scenarios, we collect the images from the Kaggle~\cite{kaggle} dataset, and use different settings to produce degradation cases.

	\noindent{\textbf{Light Transmission Disturbance:}} 
	The light transmission disturbance is achieved using Eq.~\ref{eq_d_light}. We simulate this interference using an aggregation model, which consists of global and local factors. 
	The global factors include contrast $\left(\alpha\right)$, brightness $\left(\beta\right)$ and saturation $\left(s\right)$. Thus, we simulate random color jitters with a probability of $-0.5$ to $0.5$.
	The local factors are defined as additional non-uniform illumination on the fundus image.
	To simulate the light leak phenomenon, we randomly define the panel centered at $c= (a,b) \in[0.375 r_L, 0.625 r_L]$, and $r_L\in[0.75 w, w]$, where $w$ denotes the image size.
	The Gaussian filter $G_L$ is then applied to ensure luminance smoothness. 
	Here we define $\sigma_L\in[0.66 c r_L, 0.66 (w-c) r_L]$.
	Specifically, to model the typical light leak, we define three kinds of illumination bias as $[0.63, 0.80,0.35]$, $[0.56,0.93,0.93]$, and $[1,1,1]$.
	With this, we effectively simulate the real light leak in the imaging procedure.
	Similar to the degradation model for light leak, for the uneven exposure problem, we further define $r_L\in[0.3w,0.5w]$, and the illumination bias for the underexposure model is simulated as  local brightness jitters with a probability of $-0.5$ to $-0.1$. 
	The $\sigma_L\in[0.55r_L,0.75r_L]$ of the Gaussian filter is defined for the following smooth operator.

	\noindent{\textbf{Image Blurring:}} 
	The image blurring is simulated by using Eq.~\ref{eq_d_blur}, which is caused by an undesired object distance setting during funduscopy.
	Here we set $\sigma_B=0.03 w$, and $r_B\in[0.01 w,0.015 w]$.

	\noindent{\textbf{Retinal Artifact:}}
	The retinal artifacts are achieved using Eq.~\ref{eq_d_art}. We define object $i$ as having a radius $r_i\in[0.025w,0.05w]$, while the defocused imaging is defined by a Gaussian operator $G_R$ with $\sigma_i=5+0.8r_i$.
	The illumination bias in this item is defined as $1-e^{-(0.5+0.04r_i)\times(0.012r_i)}$.
	To model the imaging of undesired objects, we randomly add 10 to 25 artifacts to each fundus image, which can simulate the interference in real clinical scenarios.
	In Fig.~\ref{fig:samples}, we provide examples of images generated with  the different degradation factors. The constructed low-quality image dataset can be used to train our model for retinal fundus image correction.
	
	\begin{figure}[!t]
		\centering
		\includegraphics[width=1\linewidth]{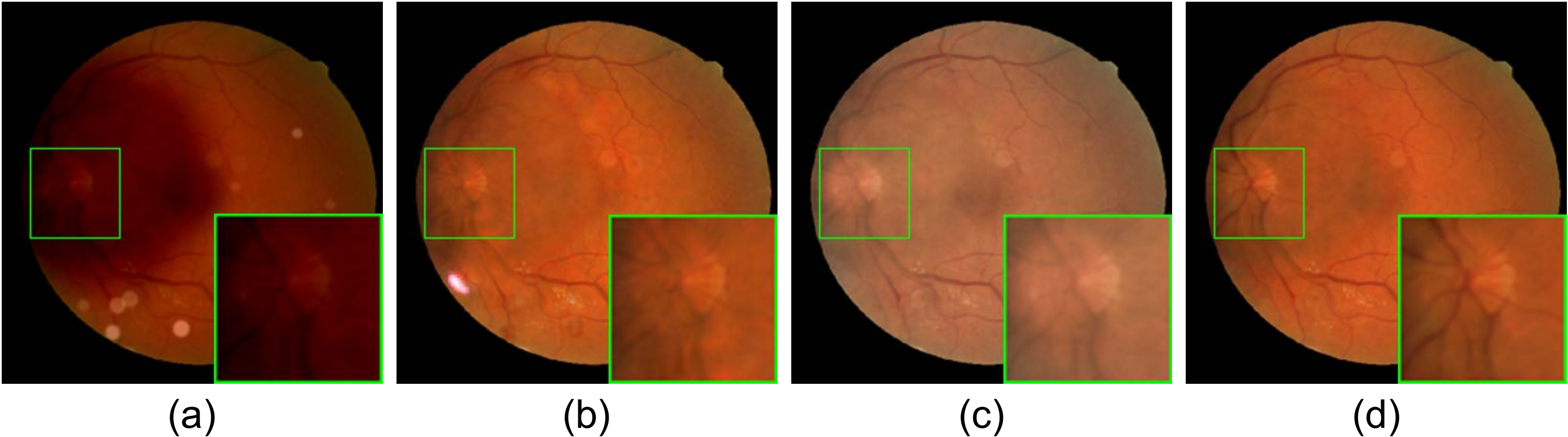}  
		\caption{\textbf{Effect of structure perceiving activation model.} (a)~Low-quality image. (b)~Backbone model. (c)~With \RSA~module. (d)~With \RSA~and \LQA~modules. The proposed method is capable of correcting the clinical fundus images with more structures and less artifacts. }\label{fig:ablation}
	\end{figure}
 
 	\begin{figure}[!t]
		\centering
		\includegraphics[width=1\linewidth]{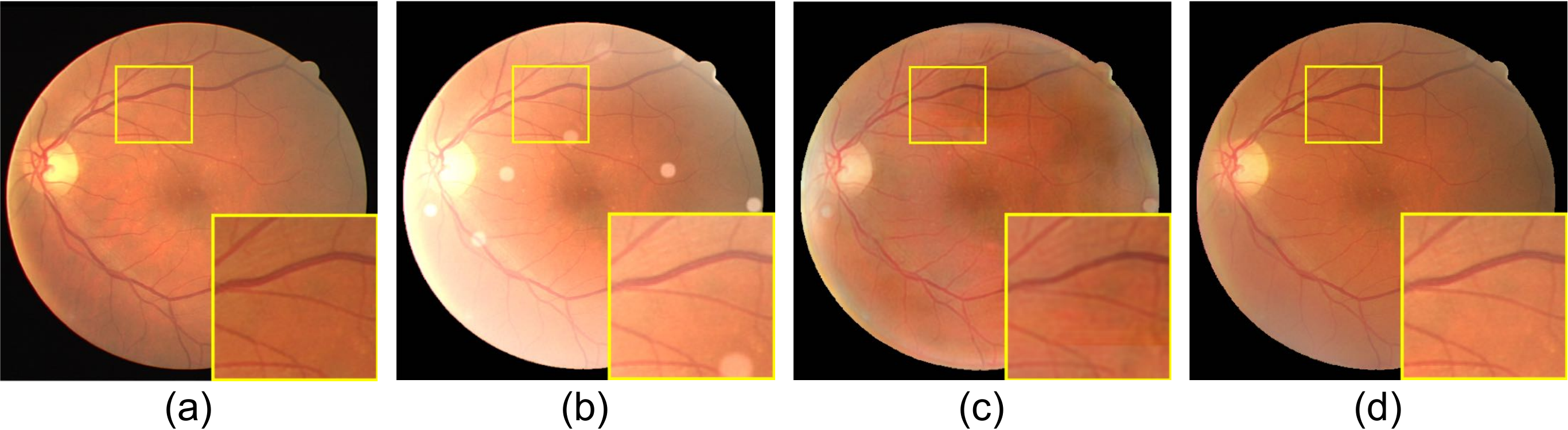}
		\caption{\textbf{Effect of multi-scale framework on enhancement.} (a)~Ground truth image. (b)~Low-quality image. (c)~Without multi-scale framework. (d)~With multi-scale framework. The proposed multi-scale framework corrects fundus images with more details. }
		\label{fig:multi}
	\end{figure}
	
	\begin{table}[!t]
		\centering
		\caption{
			\textbf{Ablation Study}.
			Our full model produces comparable results on the testing datasets in terms of various metrics.
		}
		\resizebox{0.48\textwidth}{!}{
			\setlength\tabcolsep{4.0pt}
			\renewcommand\arraystretch{1}
			\small
			\begin{tabular}{cccc||cc|cc}
				\hline 
				&&&& 
				\multicolumn{2}{c|}{DRIVE~\cite{staal2004ridge} } & \multicolumn{2}{c}{Kaggle~\cite{kaggle}} \\ 
				\multirow{-2}{*}{Multi-scale} &
				\multirow{-2}{*}{$\mathcal{L}_{\text{c}}$} &
				\multirow{-2}{*}{\RSA} &
				\multirow{-2}{*}{\LQA} &
				PSNR & SSIM & PSNR & SSIM \\
				\hline 
				&\checkmark& &  & 20.07 &0.733 & 16.86 &0.798  \\
				\checkmark&\checkmark& &  &21.06&  0.748 &17.04 &0.807 \\
				\checkmark&\checkmark&\checkmark &  &21.24&  0.740 & 18.48 &0.877  \\
				\checkmark&\checkmark&\checkmark & \checkmark &\textbf{21.24}&\textbf{0.758}&\textbf{20.51}&\textbf{0.885}  \\
				\hline 
			\end{tabular}
		}
		\label{tab:ablation}
	\end{table}

	\subsection{Ablation Study}
	To demonstrate the effectiveness of \ourmdoel, we conduct several ablation studies. 
	By removing the \RSA~and \LQA~modules from the proposed architecture, the model becomes a multi-scale baseline generation network, which is constrained by an $\mathcal{L}_2$ loss.
	As shown in Fig.~\ref{fig:ablation}~(a), unlike natural images which contain abundant mid-frequency structures and high-frequency textural details, the features represented in fundus images often appear as small vessel branches and their corresponding lesions. 
	The network optimized solely from the $\mathcal{L}_2$ loss generates relatively smooth images, as shown in Fig.~\ref{fig:ablation}~(b), preventing the tiny vessels from being reconstructed with sharp details. 
	The features and morphology of the lesions, such as hemorrhages, are also unable to be preserved.   
	To address this issue, we embed the \RSA~module as a strong prior to guide the correction procedure. 
	In Fig.~\ref{fig:ablation}~(c), our method accurately generates noticeable features e.g., vessels, and can also more accurately distinguish the disc/cup from the fundus region.
	
	Instead of simply considering generic factors affecting the imaging quality, such as  exposure and defocus, some undesired matters in initiative situations are also taken into account.
	We train the network by embedding the above architecture as well as the proposed \LQA~module.
	A trainable network is used to identify the noise region, then the specific undesired activation map is embedded with the latent features of the fundus image.
	The lateral inhibition accelerates the differentiation ability of the correction network to focus on the artifact.      
	We provide examples in Fig.~\ref{fig:ablation}~(d). As can be seen, the proposed network including both the \LQA~module and robust structure-perceiving loss function is able to better remove outlier artifacts and improve the generation performance.
	The results in Table~\ref{tab:ablation} also show that the proposed \LQA~and \RSA~modules can achieve reasonable and obvious improvements.               	

	Furthermore, to demonstrate the validity of the multi-scale framework, we also investigate the effect of removing the coarse-to-fine supervision without embedding intermediate results into the model. 
	In Fig.~\ref{fig:multi}, the fundus images contain very tiny vessels and lesions.
	Since the multi-scale network is capable of providing an incremental training strategy that encourages a robust and stable convergence procedure, it provides sharper correction results than the single-scale model.
	We also provide a quantitative comparison in Table~\ref{tab:ablation}, which demonstrates that the multi-scale framework is able to provide considerable performance promotion.

  	\begin{figure*}[!t]
		\centering
		\includegraphics[width=1\linewidth]{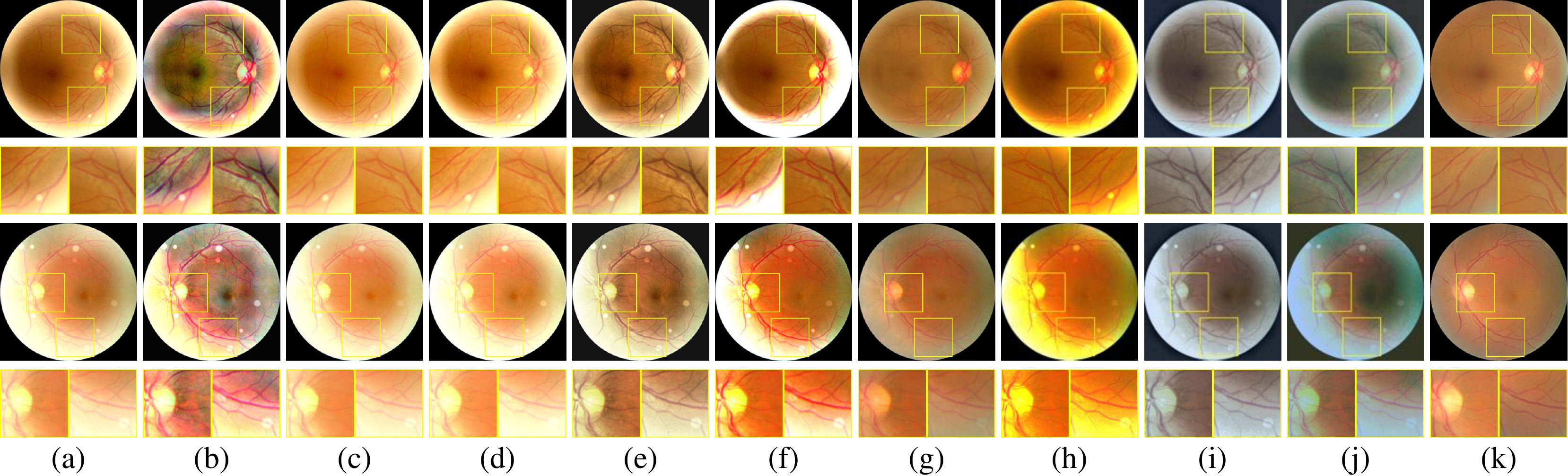}
		\caption{\textbf{Visual comparison on synthesized retinal images.} The results of proposed algorithm corrects images with preserving the clinical features and suppressing the retinal artifacts. 
		(a)~Real images. 
		(b)~Setiawan~\etal~\cite{retinaclahe}. 
		(c)~Fu~\etal~\cite{fu2016weighted}. 
		(d)~Guo~\etal~\cite{lime}. 
		(e)~Tian~\etal~\cite{naturalness2}. 
		(f)~Cheng~\etal~\cite{CHENG}. 
		(g) Eilertsen~\etal~\cite{EilertsenKDMU17}. 
		(h)~He~\etal~\cite{He0T11}. 
		(i)~Fu~\etal~\cite{convex3}. 
		(j)~Li~\etal~\cite{convex2}. 
		(k)~Ours.} 
		\label{fig:compare_sys} 
	\end{figure*}
	
    \begin{figure*}[!t]
    	\centering
    	\includegraphics[width=1\linewidth]{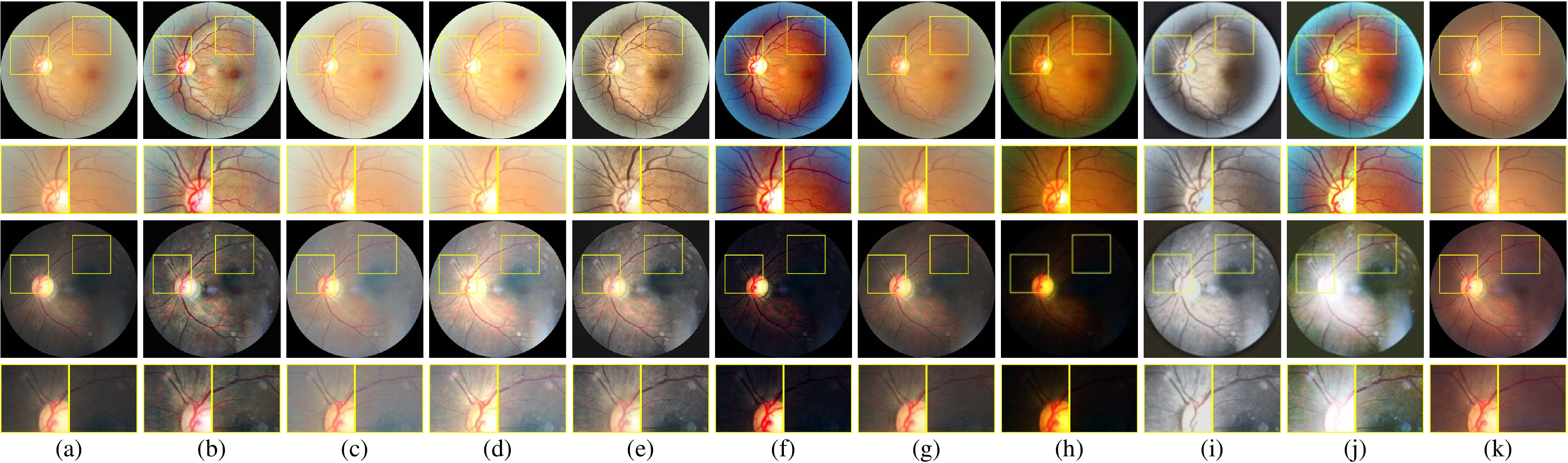}
    	\caption{\textbf{Visual comparison on real clinical retinal images.} The proposed algorithm corrects images while preserving clinical features and suppressing retinal artifacts. 
    	(a)~Real images. 
    	(b)~Setiawan~\etal~\cite{retinaclahe}. 
    	(c)~Fu~\etal~\cite{fu2016weighted}. 
    	(d)~Guo~\etal~\cite{lime}. 
    	(e)~Tian~\etal~\cite{naturalness2}. 
    	(f)~Cheng~\etal~\cite{CHENG}. 
    	(g) Eilertsen~\etal~\cite{EilertsenKDMU17}. 
    	(h)~He~\etal~\cite{He0T11}. 
    	(i)~Fu~\etal~\cite{convex3}. 
    	(j)~Li~\etal~\cite{convex2}. 
    	(k)~Ours.} 
    	\label{fig:compare} 
    \end{figure*}

	\subsection{Comparison with State-of-the-Art Methods} 
	
	In this section, we provide qualitative and quantitative comparisons with \textbf{nine} state-of-the-art  methods~\cite{zuiderveld1994,lime,fu2016weighted,naturalness2,CHENG,EilertsenKDMU17,He0T11,convex3,convex2} to demonstrate the advantages of the proposed \ourmdoel.
	We first conduct an experiment on the degraded fundus images for quantitative analysis. 
	Then, to demonstrate that our algorithm is also equally applicable to real clinical images, these methods are also used in real clinical image correction tasks.  
	As shown in Fig.~\ref{fig:compare_sys} and Table~\ref{tab:exgo}, the proposed \ourmdoel~achieves comparable performance, performing particularly favorably against the existing methods on the images with multiple degradation.
	We also provide a visual comparison on real images in Fig.~\ref{fig:compare}.
	The proposed approach effectively corrects the fundus images with relatively sharp and clean details.
	We observe that the methods~\cite{CHENG,lime,zuiderveld1994} fail to remove the noticeable undesired retinal artifacts.  
	Specifically, medical image correction aims to restore a high-quality image by suppressing noise while at the same time preserving essential pathological characteristics.    
	The handcrafted feature-based methods only consider the general image characteristics, which easily results in ignoring specific artifacts in medical images.
	In contrast, our method is capable of enhancing fundus images as well as removing these additional artifacts, which accelerates the subsequent diagnosis task.  
	
 	\begin{table}[!t]
		\footnotesize
		\centering	
		\caption{\textbf{Quantitative comparison with state-of-the-art methods.} We compute the average PSNR and SSIM on two test sets.}
		\resizebox{0.48\textwidth}{!}{
			\setlength\tabcolsep{8.0pt}
			\renewcommand\arraystretch{1}
			\small
			\begin{tabular}{c||cc|cc}
				\hline  	
				&\multicolumn{2}{c|}{DRIVE~\cite{staal2004ridge}}
				&\multicolumn{2}{c}{Kaggle~\cite{kaggle}}\\ 
				\multirow{-2}{*}{Dataset} 
				&PSNR&SSIM&PSNR&SSIM\\
				\hline
				Cheng~\etal~\cite{CHENG}&14.97&0.648&15.02&0.845\\
				Guo~\etal~\cite{lime}&14.10&0.703&13.54&0.868\\
				Fu~\etal~\cite{fu2016weighted}&15.56&0.722&14.66&0.882\\
				Tian~\etal~\cite{naturalness2}&15.42&0.721&14.71&0.664\\
				Zuiderveld~\cite{zuiderveld1994}&15.93&0.740&14.05&0.716\\
				Eilertsen~\etal~\cite{EilertsenKDMU17}&19.01&0.755&18.40&0.841\\
				He~\etal~\cite{He0T11}&15.78&0.559&15.56&0.749\\
				Fu~\etal~\cite{convex2}&10.19&0.580&9.76&0.564\\
				Li~\etal~\cite{convex3}&9.51&0.543&9.47&0.547\\
				Our \ourmdoel &\textbf{21.24}&\textbf{0.758}&\textbf{20.51}&\textbf{0.885}\\
				\hline
			\end{tabular}
		}	
		\label{tab:exgo}
	\end{table}
	
	We further provide comparisons on real fundus images, employing user studies to quantitatively evaluate the state-of-the-art methods and our method. 
	We use a paired comparison strategy to evaluate the quality of the medical images. 
	For each test, we provide 200 pairs of fundus images corrected using  different enhancement methods. We display these results in random order and ask the participants (including ophthalmologists/clinicians, and students with previous fundus image analysis experience) to rank the results based on the instructions. 
	Note that both the suppression of artifacts and preservation of lesions are taken into account.  
	Finally, we collect 65 valid responses. 
	The percentages of votes for each method are shown in Table~\ref{tab:vote}. 
	As can be seen, the proposed fundus image enhancement method receives the most votes for best corrected results.
	
	\begin{table}[!t]
		\centering	
		\caption{\textbf{Percentage scale of enhancement method in user study.}}
		\resizebox{0.4\textwidth}{!}{
			\setlength\tabcolsep{2.0pt}
			\renewcommand\arraystretch{1}
			\begin{tabular}{c||c|c}
				\hline 
				Method& Image Quality & Lesions Quality\\
				\hline 
				Setiawan~\cite{retinaclahe}&2.17&2.74\\
				Guo~\etal~\cite{lime}&7.41&5.86\\
				Fu~\etal~\cite{fu2016weighted}&5.48&2.04\\
				Tian~\etal~\cite{naturalness2}&8.12&16.88\\
				Cheng~\etal~\cite{CHENG}&9.44&11.93\\
				Eilertsen~\etal~\cite{EilertsenKDMU17}&21.42&14.82\\
				He~\etal~\cite{He0T11}&1.73&1.16\\
				Fu~\etal~\cite{convex3}&2.25&2.58\\
				Li~\etal~\cite{convex2}&1.53&2.17\\
				Our \ourmdoel &\textbf{40.45}&\textbf{39.82}\\
				\hline
			\end{tabular}
		}
		\label{tab:vote}
	\end{table}

	\begin{figure*}[!t]
		\centering
		\includegraphics[width=1\linewidth]{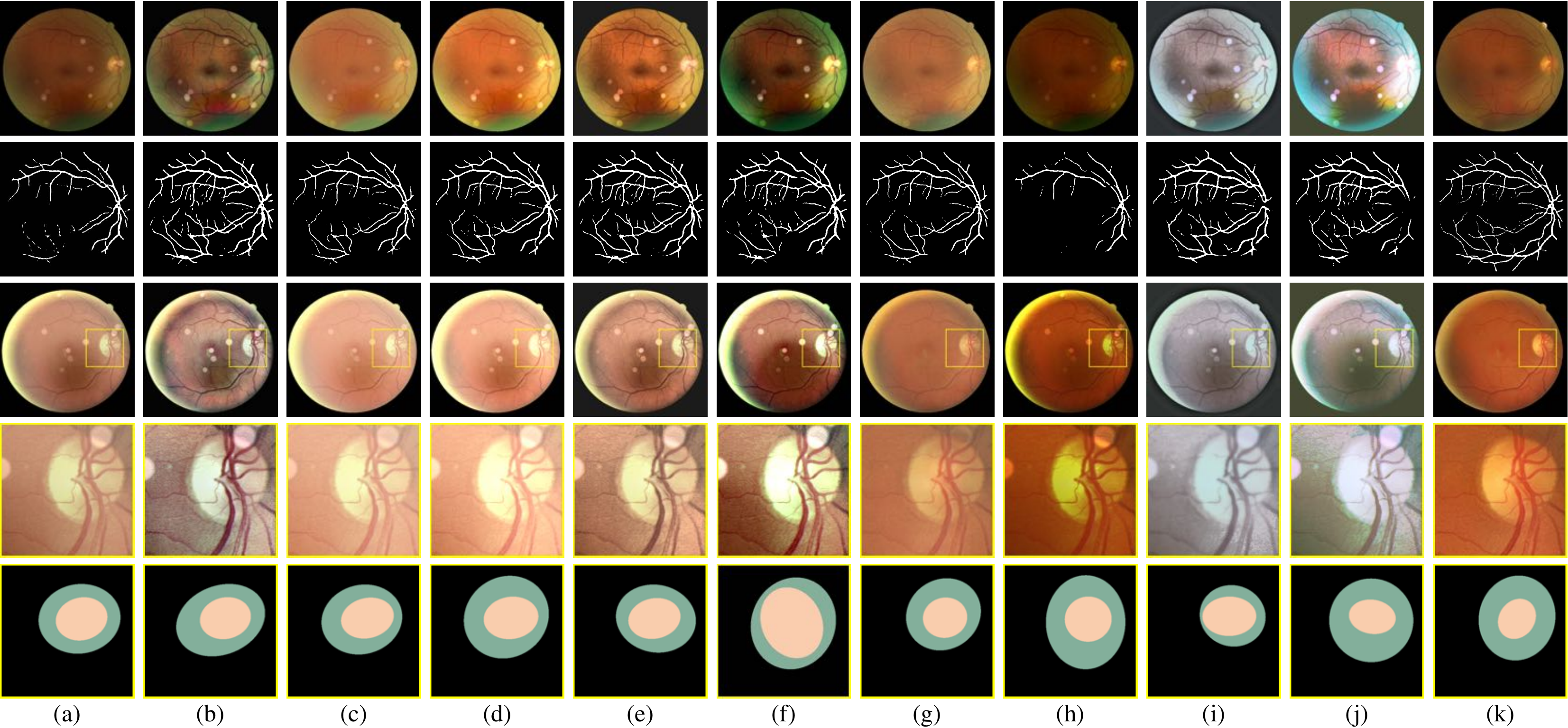}
		\caption{\textbf{Visual comparison of vessel segmentation on synthesized images.} The proposed method can improve the clinical analysis on vessel segmentation and disc/cup detection. (a)~Low quality images. (b)~Setiawan~\etal~\cite{retinaclahe}. (c)~Fu~\etal~\cite{fu2016weighted}. (d)~Guo~\etal~\cite{lime}. (e)~Tian~\etal~\cite{naturalness2}. (f)~Cheng~\etal~\cite{CHENG}. (g) Eilertsen~\etal~\cite{EilertsenKDMU17}. (h)~He~\etal~\cite{He0T11}. (i)~Fu~\etal~\cite{convex3}. (j)~Li~\etal~\cite{convex2}. (k)~Ours. }\label{fig:vessel}
	\end{figure*} 
	
    \begin{figure*}[!t]
    	\centering
    	\includegraphics[width=1\linewidth]{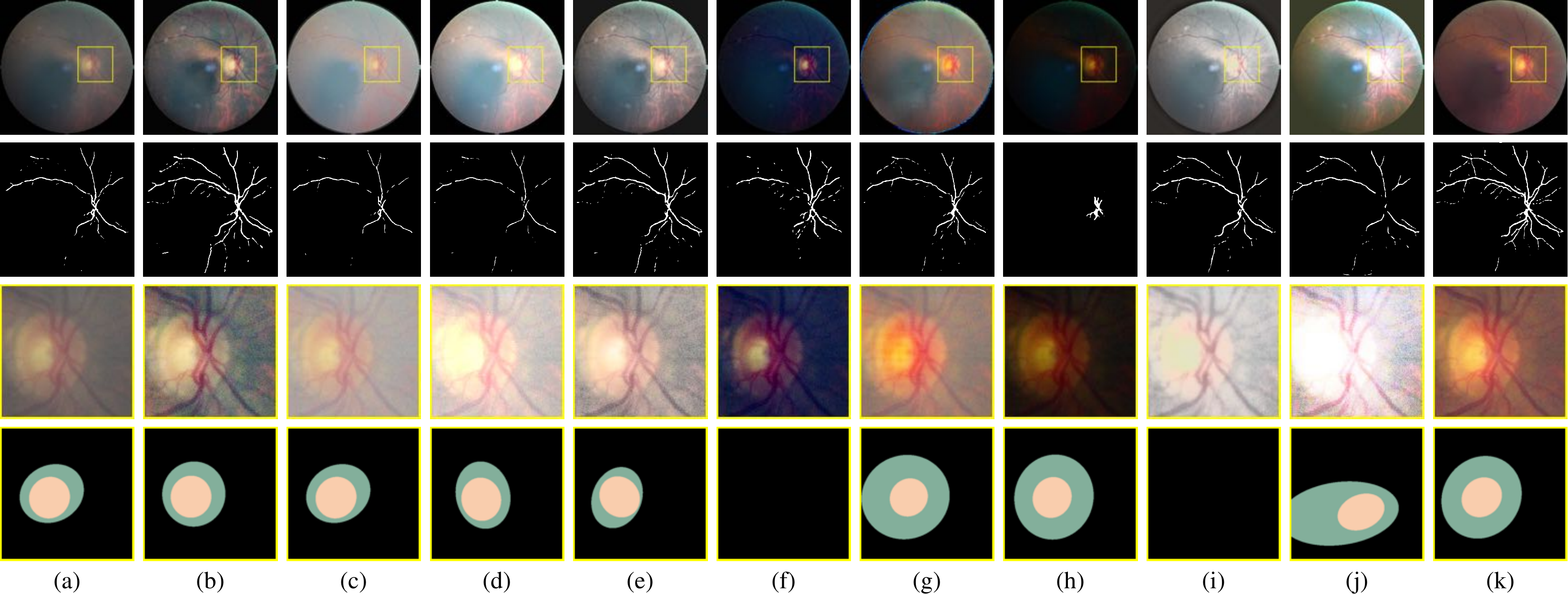}
    	\caption{\textbf{Vessel segmentation and optic disc/cup detection on real images.} (a)~Low-quality images. (b)~Setiawan~\etal~\cite{retinaclahe}. (c)~Fu~\etal~\cite{fu2016weighted}. (d)~Guo~\etal~\cite{lime}. (e)~Tian~\etal~\cite{naturalness2}. (f)~Cheng~\etal~\cite{CHENG}. (g) Eilertsen~\etal~\cite{EilertsenKDMU17}. (h)~He~\etal~\cite{He0T11}. (i)~Fu~\etal~\cite{convex3}. (j)~Li~\etal~\cite{convex2}. (k)~Ours.}
    	\label{fig:seg} 
    \end{figure*}

    \subsection{Clinical Image Analysis and Applications}
	
	Since medical image correction models should be applied to real clinical tasks, to demonstrate the effectiveness of the proposed method, we conduct additional experiments on clinical image analysis tasks, including the vessel segmentation and optic disc/cup detection.
    The DRIVE~\cite{staal2004ridge} and REFUGE~\cite{refuge} datasets are selected for evaluation, which provide the annotations of vessel and disc/cup regions. CE-Net~\cite{vessel_seg} and M-Net~\cite{disfy18} are employed as segmentation baselines.

    \begin{table}[!t]
		\centering	
		\caption{\textbf{Vessel segmentation and Disc/Cup detection evaluation}.}
		\resizebox{1\linewidth}{!}{ 
			\begin{tabular}{l||ccc|cc} 
				\hline 
				&\multicolumn{3}{c|}{Vessel Seg.}
				&\multicolumn{2}{c}{Disc/Cup Seg.}\\  
				&AUC~&Acc~&Sen~&F-score&J-score\\
				\hline 
				Without enhance								&0.924 &0.943&0.532&0.735&0.746\\
				Cheng~\etal~\cite{CHENG}				&0.950&0.940&0.586&0.713&0.705\\
				Guo~\etal~\cite{lime}					&0.953&0.951&0.541&0.829&0.835\\
				Fu~\etal~\cite{fu2016weighted}			&0.954&0.952&0.561&0.845&0.852\\
				Tian~\etal~\cite{naturalness2}			&0.941&0.945&0.575&0.822&0.803\\
				Setiawan~\cite{retinaclahe}				&0.957&0.943&0.591&0.861&0.858\\
				Eilertsen~\etal~\cite{EilertsenKDMU17}	&0.954&0.950&0.585&0.884&0.852\\
				He~\etal~\cite{He0T11}					&0.955&0.949&0.521&0.872&0.861\\
				Fu~\etal~\cite{convex3}					&0.943&0.945&0.524&0.724&0.701\\
				Li~\etal~\cite{convex2}					&0.940&0.945&0.534&0.839&0.796\\
				with enhance &\textbf{0.957}&\textbf{0.951}&\textbf{0.600}&\textbf{0.890}&\textbf{0.863}\\ 
				\hline
			\end{tabular}
		} 
		\label{tab:app}
	\end{table}
	 
	\noindent{\textbf{Vessel Segmentation:}} For the vessel segmentation task, the images  from DRIVE~\cite{staal2004ridge} are used to generate 100 low-quality images for quantitative assessment. The vessel segmentation results are shown in Fig.~\ref{fig:vessel} and Table~\ref{tab:app}. As can be seen, CE-Net fails to obtain high performance on  low-quality clinical images. In contrast, a strong vessel structure can be extracted after applying the proposed correction method. It obtains better performance on corrected images.	
	
	\noindent{\textbf{Optic Disc/Cup Detection:}} We also conduct an experiment on disc and cup detection.
	We simulate 400 degraded images with different settings from the REFUGE test set~\cite{refuge}. M-Net~\cite{disfy18} is used to segment the optic disc/cup. We report the F-scores and J-scores in Table~\ref{tab:app} and quantitative results in Fig.~\ref{fig:vessel}. Our method recovers a high dynamic range in the low-quality images, boosting the discriminative representations of the disc/cup.

    \begin{figure}[!t] 
		\centering
		\includegraphics[width=1\linewidth]{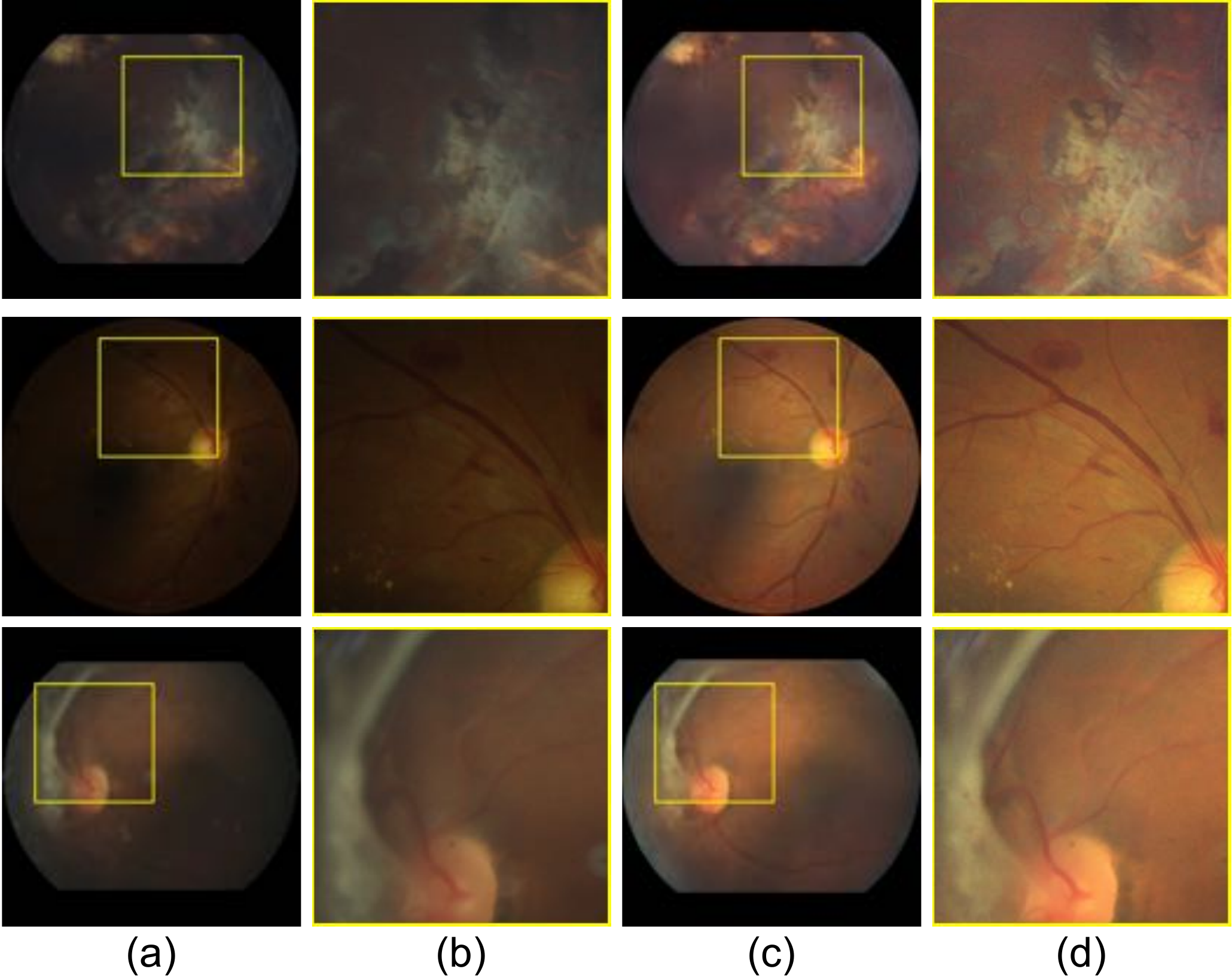} 
		\caption{\textbf{Lesion analysis on real clinical images.}  (a) Real image without correction. (b) Zoom-in of (a). (c) The image with our correction. (d) Zoom-in of (c). Our proposed method is capable of correcting images and preserving the lesions for disease analysis.}\label{fig:lesion}
	\end{figure}
	
	\noindent{\textbf{Real Clinical Fundus Image:}} To demonstrate its clinical value, we also validate the proposed correction method for vessel segmentation and optic disc/cup detection on real fundus images.
	As shown in Fig.~\ref{fig:seg}, our method produces legitimate corrected real images with clear clinical structures, which can successfully be processed by CE-Net~\cite{vessel_seg}. In addition, the optic disc/cup regions are also enhanced to be more cognizable.

    \noindent{\textbf{Real Lesion Analysis:}} To further analyze the proposed model on small lesions, we evaluate it on enhanced images from the Kaggle~\cite{kaggle} dataset. We design the degradation framework to model the artificial interference caused by ophthalmoscope imaging procedures. Our algorithm can discriminate between undesired disturbances and specific lesions. It is capable of correcting images and preserving the lesions for subsequent disease diagnosis. In Fig.~\ref{fig:lesion}, we further show the enhanced results of retinal fundus images with some lesions, such as hemorrhages and cataracts. Our method is robust to various diseases and retains the lesions, affecting the transparency of the eye's lens. 
	
	\section{Conclusion}
	In this paper, we have proposed a clinically oriented fundus enhancement network, named \ourmdoel, to correct low-quality fundus image while preserving accurate lesion areas and retinal structures. Furthermore, a complete degradation model has also been introduced to generate adequate training image pairs. Experiments support our insight into the problems of fundus image correction and degradation factor modeling. Our~\ourmdoel~can boost the performance for different clinical tasks, such as vessel segmentation and disc/cup detection. Our method can also assist ophthalmologists in ocular disease diagnosis through retinal fundus image observation and analysis, while also being beneficial to automated image analysis systems.

\bibliographystyle{IEEEtran} 
\bibliography{cofe-net}
	
\end{document}